\documentclass[
11pt,
showpacs, showkeys,
nofootinbib,
 amsmath,amssymb,
 aps,
prl,
floatfix,
]{revtex4-1}

\pdfoutput=1

\usepackage{ae,aecompl}
\usepackage[T1]{fontenc}

\usepackage{graphicx}

\usepackage{dcolumn}
\usepackage{bm}
\usepackage{hyperref}

\usepackage{subfigure}
\usepackage{longtable}
\usepackage{lipsum}
\usepackage{color}
\usepackage{comment}

\usepackage{amsmath}
\usepackage{amssymb}

\usepackage{slashbox}

\global\long\def\V#1{\boldsymbol{#1}}

\global\long\def\d#1{\delta#1}

\newcommand{\deleted}[1]{}

\newcommand{\dsum}{{\delta n}}
\newcommand{\dif}{{\delta c}}

\begin{document}

\title{Fluctuation-enhanced electric conductivity in electrolyte solutions}

\author{Jean-Philippe P\'eraud$^{1}$, Andy Nonaka$^{1}$,
John B. Bell$^{1}$, Aleksandar Donev$^{2}$, and Alejandro L. Garcia$^{3}$ }
\affiliation{$^1$ Computational Research Division, Lawrence Berkeley National Laboratory \\
 1 Cyclotron Road, Berkeley, CA 94720 \\ }
\affiliation{$^2$ Courant Institute of Mathematical Sciences, New York University \\
 251 Mercer Street, New York, NY 10012 \\ }
\affiliation{$^3$ Department of Physics and Astronomy, San Jose State University \\
 1 Washington Square, San Jose, CA 95192 \\ }

\date{\today}

\begin{abstract}
In this letter
we analyze the effects of an externally applied electric field on thermal fluctuations for a fluid containing charged species.
We show in particular that the fluctuating Poisson-Nernst-Planck equations for charged multispecies diffusion coupled 
with the fluctuating fluid momentum equation,
result in enhanced charge transport. Although this transport is advective in nature,
it can macroscopically be
represented as 
electrodiffusion with renormalized electric conductivity.
We calculate the renormalized electric conductivity by deriving and integrating the structure factor coefficients of the fluctuating quantities and show that the renormalized electric conductivity and diffusion coefficients are consistent although they originate from different noise terms. In addition, the fluctuating hydrodynamics approach recovers the electrophoretic and relaxation corrections obtained by Debye-Huckel-Onsager theory, and provides a quantitative theory that predicts a non-zero cross-diffusion Maxwell-Stefan coefficient
that agrees well with experimental measurements.
Finally, we show that strong applied electric fields result in anisotropically enhanced velocity fluctuations and reduced fluctuations of salt concentrations.

\end{abstract}



\maketitle

\paragraph*{Introduction --}
The interaction between ionic species and an externally imposed electric field is at the
core of many electrokinetic problems and applications \cite{grodzinsky2011field} such as electrophoresis. 
Studying these types of problems usually involves solving the Poisson-Nersnt-Planck equation, 
which assumes that the solution is ideal with no cross-diffusion between the different ions.
In a recent publication \cite{LowMachElectrolytes}, we presented a numerical scheme based on fluctuating hydrodynamics for simulating electrokinetic problems at mesoscopic scales where thermal fluctuations are non-negligible. In this approach, the generalized Poisson-Nernst-Planck equation is combined with the fluctuating Landau-Lifshitz Navier-Stokes equations yielding a set of stochastic partial differential equations that can be solved either analytically or numerically.
In this letter we use theoretical calculations to show that, in dilute electrolyte solutions, under an applied electric field there exists a coupling phenomenon between the fluctuations of local net charges and fluid velocity. 
This coupling results in an effective enhancement of the electric conductivity, which we call ``fluctuation-induced electroconvection''. We highlight both the similarities and the differences between this result and the enhancement of mass diffusion \cite{ExtraDiffusion_Vailati,DiffusionRenormalization_PRL,DiffusionRenormalization} associated with giant fluctuations \cite{GiantFluctuations_Nature,FluctHydroNonEq_Book}. Furthermore, we show that in the presence of an electric current there exists a coupling between the fluctuations of ion density and charge density that results in a reduction of the electric conductivity. We show that the renormalized conductivity is consistent with Onsager's reciprocal relations provided that there exists a Maxwell-Stefan (MS) cross-diffusion coefficient between the cation and anion, as, indeed, is measured in experiments.
Lastly, we show that the coupling produces an anisotropic enhancement of the momentum fluctuations of the fluid that goes as the square of the magnitude of the applied electric field.

\paragraph*{Problem description --}

We model a homogeneous solution composed of a neutral solvent fluid (e.g., water) and two ionic solute species of opposite charge. We assume that a uniform electric field $\V{E}_{\text{ext}}$ is externally applied.
Both for simplicity and for the sake of focusing on the coupling between charge fluctuations and the applied electric field, we assign the same physical parameters to the anion and the cation (i.e., equal ``bare'' diffusivity $D_0$ in the solvent, absolute charge per mass $z$, and molecular mass $m$). 
Generalizations to different ions is straightforward.
We denote the mass fraction of the cation and anion by $w_+$ and $w_-$, respectively, which are both $w_0$ in the homogeneous system. The density $\rho$, the kinematic and dynamic viscosity $\nu$ and $\eta$ ($\eta = \rho \nu$), and the dielectric permittivity $\epsilon$ are all assumed constant. We assume that the system remains isothermal at temperature $T$ and neglect both viscous and ohmic heating.

The theoretical system we consider is infinite in all directions. 
The fluid is subjected to fluctuations in species mass flux and stress tensor consistent with the fluctuation-dissipation theorem \cite{FluctHydroNonEq_Book}.
We use a low Mach approximation \cite{LowMachExplicit} and neglect density fluctuations, i.e., $\V{\nabla} \cdot \V{v}=0$, where $\V{v}(\mathbf{x},t)$ refers to the velocity vector with components $(v_x, v_y, v_z)$.
Assuming the electrolytes are dilute, the equations describing the mass fractions are:
\begin{equation}
\partial_t w_{\pm} + \mathbf{v} \cdot \nabla w_{\pm} = D_0 \nabla^2 w_{\pm} \mp \frac{D_0 m z}{k_B T} \nabla \cdot \left( \V{E} w_{\pm} \right)  + \nabla \cdot \left(\sqrt{2 D_0 m {\rho}^{-1} w_\pm}  \V{Z}_\pm \right)
\label{mass_conservation}
\end{equation}
where $k_B$ is Boltzmann's constant. We used Nernst-Einstein relation, which states that the electric mobility is given by $D_0 m z/(k_B T)$. Symbols $\V{Z}_+$ and $\V{Z}_-$ refer to two independant Gaussian white noise vector fields, and, assuming that the dielectric coefficient is constant, the total electric field is the solution to $\epsilon \nabla \cdot \V{E}=z \rho (w_+-w_-) \equiv q_f$.
The velocity field follows the fluctuating Navier-Stokes equation
\begin{equation}
\partial_t \V{v} + \nabla \cdot \left( \V{v} \V{v}^T \right) = \nu \nabla^2  \V{v} + {\rho}^{-1}{\nabla p}  + {\rho}^{-1}{q_f\V{E}}+\sqrt{\nu {\rho}^{-1} k_B T}\;\nabla \cdot \left( \V{\mathcal{W}}+\V{\mathcal{W}^T} \right)
\label{momentum}
\end{equation}
where $p$ is the pressure and $\V{\mathcal{W}}(\V{r},t)$ is a white noise tensor field; superscript $T$ denotes transpose.
Physically, the fluctuation-induced electroconvection we are studying is due to mass fraction fluctuations, given by Eq. \eqref{mass_conservation}, which result in enhanced velocity fluctuations through the term $q_f\V{E}$ in Eq. \eqref{momentum}. The calculations we carry out next closely resemble previous linearized or ``one-loop renormalization'' calculations of fluctuation-enhanced diffusivity in non-ionic binary mixtures \cite{ExtraDiffusion_Vailati,DiffusionRenormalization_PRL,DiffusionRenormalization}.

\paragraph*{Structure factors --}
We now use linearized fluctuating hydrodynamics to compute the spectrum of the steady-state concentration and velocity fluctuations. We define the fluctuations $\delta w_\pm=w_{\pm}-w_0$ but use the sum $\dsum = \delta w_++\delta w_-$ and difference $\dif=\delta w_+-\delta w_-$, which are more suited to describe the problem than the individual mass fractions. Linearizing Eq. \eqref{mass_conservation} yields:
\begin{equation}
\begin{split}
\partial_t \dsum  &= D_0 \nabla^2 \dsum  - \frac{D_0 m z}{k_B T} \V{E}_{\text{ext}} \cdot \nabla \dif + \sqrt{2 D_0 m {\rho}^{-1} w_0}\nabla \cdot ( \V{Z}_+ +  \V{Z}_-) \\
\partial_t \dif  &= D_0 \nabla^2 \dif - \frac{D_0}{\lambda^2} \dif  - \frac{D_0 m z}{k_B T} \V{E}_{\text{ext}} \cdot \nabla \dsum + \sqrt{2 D_0 m {\rho}^{-1} w_0}\nabla \cdot ( \V{Z}_+- \V{Z}_- )
\end{split}
\label{lin_mass_conservation}
\end{equation}
where the Debye length $\lambda$ is defined by $\lambda^2=\epsilon k_BT/(2\rho mw_0 z^2)$.
We also linearize \eqref{momentum} and, as in \cite{FluctHydroNonEq_Book}, we apply a double curl operator in order to eliminate the pressure term.
We obtain, in Fourier space:
\begin{equation}
\begin{split}
-\V{k}\times \V{k} \times& \partial_t \hat{\V{v}}=-\nu k^4 \hat{\V{v}}-z\V{k} \times \V{k} \times \left [ \V{E}_{\text{ext}}\hat{\dif}\right ] \\ 
                                                                 &-i \sqrt{\nu {\rho}^{-1} k_B T} \V{k} \times \V{k} \times \left[ \V{k} \cdot (\hat{\mathcal{W}} + \hat{\mathcal{W}}^T) \right ]
\end{split}
\label{momentum_fourier}
\end{equation}
We take $\V{E}_{\text{ext}}=E_{\text{ext}} \V{e}_x$, where $\V{e}_x$ is a unit vector in the $x$ direction, and let $\theta$ denote the angle between $\V{E}_{\text{ext}}$ and the wavevector $\V{k}$. In that case, the $x-$component of \eqref{momentum_fourier} becomes:
\begin{equation}
\partial_t \hat{v_x} = -\nu k^2 \hat{v_x} + zE_{\text{ext}}\sin^2(\theta)\hat{\dif} + ik \sin(\theta) \sqrt{2\nu {\rho}^{-1} k_B T} \hat{\mathcal{V}}
\label{momentum_fourier_short}
\end{equation}
and where $\hat{\mathcal{V}}(\V{k},t)$ is a scalar white-noise process.

Taking the Fourier transform of \eqref{lin_mass_conservation} and combining it with \eqref{momentum_fourier_short}, we obtain that the vector $\hat{\mathcal{U}}(\V{k},t) =( \hat{\dsum}(\V{k},t), \hat{\dif}(\V{k},t), \hat{v}_x(\V{k},t))$ is described by the Ornstein-Uhlenbeck process $\partial_t \hat{ \mathcal{U} }=\V{M} \hat{\mathcal{U}} + \V{N} \hat{\mathcal{Z}}$
with
\begin{equation}
\V{M}=\left ( \begin{array}{c c c}
-D_0k^2  & -i k \cos(\theta)\frac{E_{\text{ext}}D_0 m z}{k_B T} & 0 \\
 -i k \cos(\theta)\frac{E_{\text{ext}}D_0 m z}{k_B T}  & -D_0\left(k^2 +\lambda^{-2} \right)  & 0 \\
0 & zE_{\text{ext}} \sin^2(\theta)& -\nu k^2
\end{array} \right )
\label{relaxation_matrix}
\end{equation}
where $\hat{\mathcal{Z}}(\V{k},t)$ is a vector of three uncorrelated white noise processes.
The variance matrix is diagonal,
\begin{equation}
\V{N}\V{N}^* = k^2 \rho^{-1} \text{Diag}\left\{ 4mD_0 w_0, 4mD_0 w_0, 2\nu k_B T \sin^2( \theta) \right\}.
\nonumber
\end{equation}
The steady-state spectrum of the fluctuations, i.e., the matrix of static structure factors $\V{S}(\V{k})=\langle \hat{\mathcal{U}} \hat{\mathcal{U}}^*\rangle$, where $\langle \cdot \rangle$ denotes the steady-state average, is given by the solution of the linear system $\V{M} \V{S} + \V{S} \V{M}^* = -\V{N}\V{N}^*$ \cite{FluctuationDissipation_Kubo}.

The complete expression for $\V{S}(\V{k})$ is quite involved. 
Here we focus on the linear response to the applied field. For sufficiently weak electric fields, there are only two correlations that are altered by the electric field to linear order in $E_{\text{ext}}$:
\begin{align}
S_{\hat{\dif},\hat{v}_x} &= \frac{2 m w_0}{\rho D_0} \frac{z k^2 \lambda^4 \sin^2( \theta )}{\left[ 1+(\text{Sc}+1)k^2 \lambda^2 \right] \left[1+ \lambda^2 k^2 \right]} E_{\text{ext}} \label{alpha_linear}\\
S_{\hat{\dif},\hat{\dsum}}&=i \frac{2 m w_0}{\rho k_B T} \frac{m z k \lambda^2 \cos( \theta )}{(1+k^2 \lambda^2)(1+2 k^2 \lambda^2)} E_{\text{ext}} \label{eq:S_cn}
\end{align}
The auto-correlations $S_{\hat{\dsum},\hat{\dsum}}$, $S_{\hat{\dif},\hat{\dif}}$ and $S_{\hat{v}_x,\hat{v}_x}$ are, to leading order, quadratic in $E_{\text{ext}}$.

\paragraph*{Enhancement of electric conductivity --}
The electroconvective coupling results in a net charge flux.
From \eqref{mass_conservation}, we may write the average charge flux as:
\begin{equation}
 \langle \V{F}_{q} \rangle =  \underbrace{\frac{2 \rho D_0 m z^2}{k_B T}\langle \V{E} \rangle w_0}_{\V{F}_{q,0}}  +  \underbrace{\rho z \langle \V{v} \dif \rangle}_{\V{F}_{q,\text{adv}}} +  \underbrace{\frac{ \rho D_0 m z^2}{k_B T}\langle \d{\V{E}} \dsum \rangle}_{\V{F}_{q,\text{relx}}},  
\end{equation}
where $\d{\V{E}}\equiv \V{E}-\langle \V{E} \rangle$.
Here $\V{F}_{q,\text{0}}=C_0 \V{E}_{\text{ext}}$ where $C_0 \equiv 2 m \rho w_0 z^2 D_0/(k_B T)$ is the electric conductivity resulting from the Nernst-Einstein relation. On the other hand the two other terms modify the charge flux because the correlations $\langle \V{v} \dif \rangle$ and $\langle \d{\V{E}} \dsum \rangle$ are non-zero as we show below. This additional charge flux is proportional to the electric field in the linearized regime and can be related to an enhanced electric conductivity. 



%

We first examine the advective charge flux $\V{F}_{\text{adv}}$, which is intuitively the most direct consequence of the coupling and results from the correlation between the velocity and the charge density fluctuations. It is also the most important quantitatively. We can physically interpret $ \lim_{k\rightarrow \infty} S_{\hat{\dif},\hat{v}_x} = 0$ as charge fluctuations with small wavelength diffusing away before the Lorentz force can advectively accelerate the charged regions.
The component of the advective flux parallel to $\V{E}_\text{ext}$ can be expressed as an integral of Fourier components over all wavevectors,
\begin{align}
\V{F}_{q,\text{adv}} \cdot \V{e}_x & =C_{\text{adv}} E_{\text{ext}} =\frac{\rho z}{8\pi^3 } \int_{k<k_c} S_{\hat{\dif}, \hat{v}_x} d \V{k} \label{F_x_int_kvec} \\
& = \frac{\rho z}{4\pi^2} \int_{k=0}^{k_c} \int_{\theta=0}^{\pi} S_{\hat{\dif},\hat{v}_x} k^2 \sin(\theta) d \theta d k  \label{F_x_int_k}
\end{align}
where, as done in prior work on renormalization of diffusion coefficients \cite{DiffusionRenormalization}, we define a cutoff $k_c= {\pi}/{a}$, where $a$ is a molecular scale. This is necessary since the integrand is not integrable because it converges towards a non-zero quantity for large wavenumbers. This ``ultraviolet divergence'' is actually a consequence of a breakdown of the validity of the hydrodynamic equations at molecular scale.
Performing the integral in \eqref{F_x_int_k} using \eqref{alpha_linear}, and using the fact that the Schmidt number in liquids is large, $\text{Sc} \gg 1$, we obtain the approximation
\begin{eqnarray}
C_{\text{adv}}& \approx& \frac{2m  w_0 z^2}{3\pi D_0 a \text{Sc}} 
\left[
1 -  \frac{a}{\pi \lambda} \arctan \left( \frac{\pi \lambda}{a} \right)
   \right]\\
\label{netAdvectiveCurrent}   
& \approx& \frac{2m \rho w_0 z^2}{k_B T} 
\left[
\frac{k_B T}{3\pi a \eta} -  \frac{k_B T}{6\pi \lambda \eta}
   \right] \equiv C_{\text{enh}}+C_{\text{ep}},    
\end{eqnarray}
where in \eqref{netAdvectiveCurrent} we expand to leading order in $a/\lambda$ since $\lambda \gg a$ for dilute solutions. We note that $C_\text{ep}$ is known as the electrophoretic term, derived within the Debye-Huckel-Onsager (DHO) theory by rather different means  ~\cite{robinson2012electrolyte, Onsager1927}.
We note that the term in bracket in \eqref{netAdvectiveCurrent} can be interpreted as a difference of Stokes-Einstein coefficients for a sphere of radius $a/2$ and a sphere of radius $\lambda$. This corresponds to the classical physical picture that the Stokes friction on an ion needs to be adjusted because an ion must drag its ionic atmosphere with it ~\cite{ElectrolytesMS_Review} (equivalently, the ion experiences fluid drag relative to ionic cloud~\cite{robinson2012electrolyte}).

The flux $\V{F}_{ \text{relx}}$ is derived here by using the fact that $\epsilon \langle \V{E} \dsum \rangle = \rho z \langle \nabla \left[ \nabla^{-2} \dif \right] \dsum \rangle $ and going to Fourier space:
\begin{equation}
\V{F}_{q,\text{relx}}  = \frac{ \rho^2 D_0 m z^3}{8 \pi^3 \epsilon k_B T  } \int_{\V{k}}i\frac{\V{k}}{k^2} S_{\hat{\dif}, \hat{\dsum}} d \V{k} \equiv  C_{\text{relx}} \V{E}_{\text{ext}}
\end{equation}
which, after using Eq. \eqref{eq:S_cn}, becomes:
\begin{equation}
C_{\text{relx}}= -\frac{D_0 \rho m^3 z^4 w_0 \sqrt{2}}{12 \lambda \pi \epsilon k_B^2 T^2 (1+\sqrt{2})}. 
\label{relaxationCurrent}
\end{equation}
Physically, this is due to the anisotropic counter-ionic cloud surrounding a given ion and known in the DHO theory as the relaxation term ~\cite{robinson2012electrolyte, Onsager1927}.

Both $C_{\text{ep}}$ and $C_{\text{relx}}$ go as $w_0^{1/2}$ and vanish in the limit of infinitely dilute solutions where $\lambda \gg a$. Expressions \eqref{netAdvectiveCurrent} and \eqref{relaxationCurrent} show that the deterministic linear response that is obtained by ensemble-averaging the equations is not the ``bare'' response expressed by the conductivity $C_0$, but is instead enhanced, or {\it renormalized} by the enhanced conductivity $C_{\text{enh}}$ due to fluctuation-induced charge transport. Macroscopically, this suggests that the quantity that is experimentally accessible is the renormalized or ``dressed'' $C = C_0+C_\text{enh}+C_{\text{ep}}+C_{\text{relx}}$, and that particular care should be taken when setting the simulation parameters of a fluctuating hydrodynamics solver, so that this enhancement effect is not double-counted \cite{LowMachExplicit}. 

\paragraph*{Renormalized transport coefficients --}
The renormalization of the electric conductivity is connected to the renormalization of the diffusion coefficient that results from giant fluctuations \cite{ExtraDiffusion_Vailati,DiffusionRenormalization_PRL,DiffusionRenormalization}. 
In \cite{DiffusionRenormalization}, a calculation very similar to the one performed above is carried out for the renormalization of the diffusion coefficient in a non-ionic mixture, and it is found that diffusion is renormalized by \footnote{Quantitatively, assigning the experimental self diffusion coefficient of the ions to $D_{\text{enh}}$ and $\eta$ provides estimates of the lengthscale $a$ on the order of the ionic radii.} $D_{\text{enh}} = k_BT/(3 \pi a \eta)$. While this result was derived for non-ionic solutions, it can easily be generalized since analyzing the giant fluctuations in the linear regime requires imposing electroneutrality of the steady state. Consequently, the macroscopic gradients of the species charge densities must be equal, which in our case reduces to $\nabla w_+ = \nabla w_-= \nabla w_0$. With this condition, the approach developed in \cite{DiffusionRenormalization} shows that the renormalized mass flux for $\dsum$ is the same as that of non-ionic solutions. Qualitatively, the renormalization of the diffusion coefficient is not affected by the presence of charges because the thermal velocity fluctuations advect both the ion and the counterion together, thus maintaining electroneutrality. As with non-ionic mixtures, the renormalized diffusion coefficient is $D \equiv D_0 + D_{\text{enh}}$.
On the other hand, the electric conductivity $C=C_0+C_\text{enh}+C_{\text{ep}}+C_{\text{relx}}$ is renormalized to:
\begin{eqnarray}
C & \approx & \frac{2mw_0 z^{2}\rho}{k_{B}T}\left(D_{0}+\frac{k_{B}T}{3\pi a\eta}-\frac{k_{B}T}{6\pi\lambda \eta} - \frac{D_0 z^2 m^2}{12(2+\sqrt{2}) \pi \lambda \epsilon k_B T}\right)\label{eq:cond_renorm}\\
 & \approx & \frac{2mw_0 z^{2}\rho}{k_{B}T}\left(D-\frac{A}{\lambda}\right)=C_{\text{PNP}}-C_0\frac{A}{D_0\lambda}.\nonumber 
\end{eqnarray}
where $C_{\text{PNP}}= (2mw_0 z^{2}\rho/k_{B}T)D$ is the electric conductivity obtained from the Poisson-Nernst-Planck equations with the renormalized diffusivity $D$ and where $A$ is a coefficient independent of the concentration of electrolytes.

For infinitely dilute solutions ($\lambda \rightarrow \infty$), the renormalizations of the electric conductivity and the diffusivity are consistent with the Poisson-Nernst-Planck equation, i.e., $C=C_{\text{PNP}}$, which amounts to assuming that Fick's diffusion matrix is diagonal. This is a manifestation of the overall consistency of fluctuating hydrodynamics, even though the two enhancement phenomena stem from distinct noise terms~\footnote{The renormalization of diffusion originates from the velocity fluctuations and their coupling with a concentration gradient, while the renormalization effect studied here results from charge density fluctuations and their coupling with the electric field}; it is worth noting that in the fully nonlinear diffusion model studied in~\cite{DiffusionJSTAT} the only noise term is the stochastic stress and all diffusion arises by advection by thermal velocity fluctuations.

For finite $\lambda$, the renormalized diffusion coefficient $D_0+D_{\text{enh}}$ and the renormalized electric conductivity \eqref{eq:cond_renorm} do not satisfy the Nernst-Einstein relation so the renormalized Poisson-Nernst-Planck equation must be corrected to leading order in $a/\lambda$ to be consistent with Onsager's reciprocal relations. Specifically, the renormalized Fick's diffusion matrix must include off-diagonal coefficients; to satisfy both renormalized coefficients, the mass fluxes $\V{F}_+$ and $\V{F}_-$ of the two ionic species must be expressed as: 
\begin{equation}
\left(
\begin{array}{c}
\V{F}_+ \\ 
\V{F}_- 
\end{array}
\right)
= -\rho
\underbrace{
\left(
\begin{array}{c c}
D-\frac{A}{2 \lambda} &  \frac{A}{2 \lambda} \\
 \frac{A}{2 \lambda} & D-\frac{A}{2 \lambda} 
\end{array}
\right)
}_{\V{D}}
\cdot \nabla
\left(
\begin{array}{c}
   w_+ + \frac{m z}{k_B T} w_+ \phi \\ 
   w_- - \frac{m z}{k_B T} w_- \phi
\end{array}
\right)
\end{equation}
where $\phi$ is the electric potential ($\V{E}=-\nabla \phi$).

In order to give a more physical interpretation to the cross-diffusion coefficient, we link the renormalized Fickian diffusion matrix to a renormalized Maxwell-Stefan (MS) diffusion matrix \cite{IrrevThermoBook_Kuiken}. The MS diffusion coefficients can be physically interpreted as inverse friction coefficients between \emph{pairs} of distinct species. 
For a very dilute solution, it has been assumed when writing Eqs. \eqref{mass_conservation} that the (bare) MS cross-diffusion coefficient between the two ionic species, $\mathfrak{D}^{(+,-)}_0$, is 0, and that the (bare) cross-diffusion coefficient between the solvent and either ion is identical, i.e. $\mathfrak{D}^{(s,+)}_0=\mathfrak{D}^{(s,-)}_0=D_0$.
However, this is inconsistent with the renormalized Fickian diffusion matrix $\V{D}$ with nonzero off-diagonal coefficients. Introducing the renormalized MS diffusion coefficients $\mathfrak{D}^{(+,-)}$ and  $\mathfrak{D}^{(s,+)}=\mathfrak{D}^{(s,-)}$ and writing the friction matrix as the inverse of the Fickian diffusion matrix, we obtain, to first order in $w_0$, $ \mathfrak{D}^{(s,+)}= D$, and the cross-diffusion coefficient:
\begin{equation}
\mathfrak{D}^{(+,-)} \approx 12 \pi  D^2  \left[ \frac{k_B T}{\eta} + \frac{D_0 z^2 m^2}{2(2+\sqrt{2}) \epsilon k_B T } \right]^{-1} \frac{M}{m} \lambda w_0 
\label{cross_diffusion}
\end{equation}
where $M$ denotes the molecular mass of the solvent.

Using the complete formulas for the electrophoretic ($C_{\text{ep}}$) and relaxation ($C_{\text{relx}}$) terms from DHO theory \cite{robinson2012electrolyte}, one can easily generalize Eq. \eqref{cross_diffusion} to unequal ions. With parameters of water (molecular mass $M=3\times 10^{-26}$ kg, $\eta=1.05 \times 10^{-3}$ kg/ms), we find $\mathfrak{D}^{(+,-)} \approx 0.9 \times 10^{-10} \sqrt{c}$ for salt solutions ($D_{\text{Na}} \approx 1.3 \times 10^{-9}$ m$^2$/s and $D_{\text{Cl}} \approx 2.0 \times 10^{-9}$ m$^2$/s) where $c$ is in mol/L and where the result is in m$^2$/s, in very good agreement (within 10\% difference) with published experimental measurements \cite{ElectrolytesMS_Review,Visser_Thesis,LatticeBoltzElectrolytes}.

\begin{figure}
  \centering
  \includegraphics[width=0.48\textwidth]{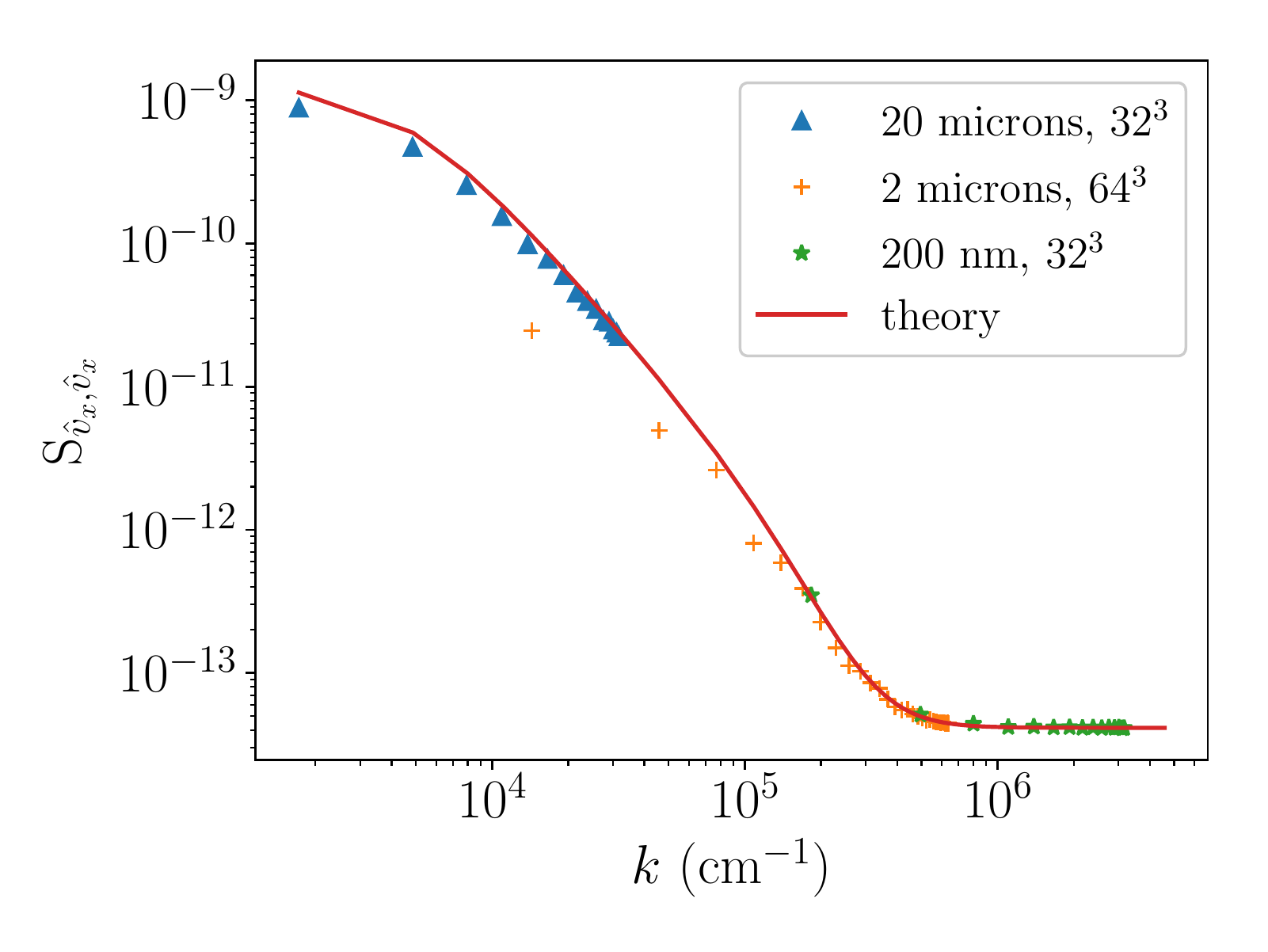}
  \caption{Structure factor of velocity fluctuations parallel to the applied field versus wavenumber. The computational system used to verify the theoretical calculations is a cubic domain of dimension $L$, with periodic boundary conditions in all directions, with: $T=300$ K, $\epsilon= 6.91 \times 10^{-19}$ s$^2\cdot$C$^2\cdot$cm$^{-3}\cdot$g$^{-1}$, $\nu=1.05\times 10^{-2}$ cm$^2 \cdot$s$^{-1}$, $D_0=10^{-5}$ cm$^2 \cdot$s$^{-1}$, $z=10^3$ C$\cdot$g$^{-1}$, $w_0=10^{-5}$, $m=3 \times 10^{-23}$ g, $\rho=1.0$ g$\cdot$ cm$^{-3}$, $E=10^{6}$ V $\cdot$ cm$^{-1}$=10$^{13}$g $\cdot$ cm$\cdot$s$^{-2}\cdot$C$^{-1}$. Since a single computation can not cover the wide range of wavenumbers shown here, the computational results combine three different systems, of sizes 20 microns, 2 microns and 200 nm. The number of computational cells is indicated in the legend. The theoretical calculation is corrected to account for the discrete Laplacian effect \cite{LowMachElectrolytes}.}
  \label{structure_factor_vx}
\end{figure}

\paragraph*{Enhancement of velocity fluctuations --}
The fluctuation-induced electroconvection derived in this paper is associated with a corollary phenomenon, namely, the enhancement of velocity fluctuations in the direction of the electric field, as shown by the expression of $S_{\hat{v}_x,\hat{v}_x}$, written below in the case where $\V{k}$ and $\V{E}_{\text{ext}}$ are orthogonal ($\theta=\pi/2$):
\begin{equation}
S_{\hat{v}_x,\hat{v}_x}^\perp = \frac{k_BT}{\rho} + \frac{2m w_0}{\rho\nu D_0}\frac{ z^2 E_{\text {ext}}^2  \lambda^4}{ \left[1+ (\text{Sc}+1) k^2 \lambda^2 \right] \left[1 + \lambda^2 k^2 \right]} \label{S_vv}
\end{equation}
In Figure \ref{structure_factor_vx} we show a comparison between the theoretical structure factor of velocity fluctuations (when the wavevector and the electric field are orthogonal) and the same quantity obtained with the code developed in \cite{LowMachElectrolytes}. The main finding here is that, provided the field is strong enough, the amplitude of the low wavenumber fluctuations is noticeably enhanced. As in the phenomenon of giant fluctuations \cite{GiantFluctuations_Nature,FluctHydroNonEq_Book}, this results in large scale patterns, with the key difference that these patterns are found in the $x-$component (i.e. colinear to the applied electric field) of the velocity instead of the mass fractions.

For small wavenumbers (large length scales), the structure factor for wavevectors orthogonal to $\V{E}_{\text{ext}}$ ($\theta = \pi/2$) converges toward
\begin{equation}
S_{\hat{v}_x,\hat{v}_x}^\perp (k \rightarrow 0) = \frac{k_BT}{\rho} \left[ 1 + \frac{\epsilon E_{\text{ext}}^2 \lambda^2}{\rho \nu D_0 } \right].
\end{equation}
The effect of the electric field on the velocity fluctuations is significant when
$E_{\text{ext}} \geq \lambda^{-1} \sqrt{ \epsilon^{-1} \rho \nu D_0}$.
Using the Maxwell approximation $\nu\approx v_{\text{th}}\lambda_{\text{th}}$ where $v_{\text{th}}$ and $\lambda_{\text{th}}$ refer respectively to the molecular speed and length (e.g., sound speed and mean free path), we can write it as
\begin{equation}
\frac{\epsilon E_{\text{ext}}^2/2}{\rho v_{\text{th}}^2/2} \gtrsim \frac{1}{\text{Sc}} \frac{\lambda_{\text{th}}^2}{\lambda^2},
\end{equation}
where the left-hand side is the ratio of the electric and the thermal energy densities. This condition may seem constraining, but in dilute liquid solutions the right hand side is much smaller than 1. In fact, with the parameters chosen for Figure \ref{structure_factor_vx}, the condition on the electric field is $E_{\text{ext}} \gtrsim$ 6 kV/cm which is the higher end of the electric fields applied in electrophoresis experiments \cite{Verpoorte2002}.

We note that, on the other hand, the fluctuations of $\dsum$ are reduced anisotropically by the electric field,
\begin{align}
S_{\hat{\dsum},\hat{\dsum}} (k \rightarrow 0) &=
2 m w_0 \rho^{-1} \left[1+\left(\frac{mzE_{\text{ext}}\cos(\theta)\lambda}{k_B T} \right)^2 \right]^{-1} \\
&= 2 m w_0 \rho^{-1} \left[1+ \frac{m \epsilon E_{\text{ext}}^2 \cos^2(\theta)}{\rho n_0 k_B T} \right]^{-1},
\end{align}
where $n_0$ is the total ion mass fraction.
Note that the second term in the brackets is the ratio of the typical magnitude of the Maxwell stress tensor and the osmotic pressure of the ions.
The reduction of the ion number density fluctuations is significant when $m z \lambda E_{\text{ext}} \gtrsim k_B T$, or, equivalently, when the energy lost (or gained) by an ion crossing a distance $\lambda$ in the direction of the field is larger than the thermal energy $k_B T$. Using parameters for sodium at concentration $w_0=10^{-5}$ gives $E_{\text{ext}} \gtrsim $ 20 kV/cm.
\paragraph*{Concluding remarks --}
In summary, using a fluctuating hydrodynamics formulation, we show that there exists a coupling between the fluctuations in charge density and fluid velocity that is proportional to the applied electric field. This coupling leads to an effective enhancement, or renormalization, of the measured electric conductivity of an ionic mixture. This enhancement is comparable to the enhancement of the diffusion coefficients that results from giant fluctuations, in that the enhancement coefficients match in the limit of infinite dilution. For finite dilution, the renormalization of mass diffusivity and electric conductivity are different. This shows that, although we started from a diagonal Fickian diffusivity matrix, renormalizing the fluctuating Poisson-Nernst-Planck equations yields an off-diagonal Fickian diffusion term, itself linked with a non-zero renormalized cross-diffusion Maxwell-Stefan coefficient between the two counterions, in good agreement with experimental coefficients reported in the literature. In fact, in our prior work \cite{LowMachElectrolytes} 
we demonstrated that results from Debye-Huckel theory, including the non-analytic
Debye-Huckel correction to the internal energy, can be obtained from a fluctuating
hydrodynamics theory of dilute electrolyte solutions. The present work further demonstrates that fluctuating hydrodynamics provides a generalizable and systematic approach to derive corrective transport coefficients such as the electrophoretic and the relaxation term. Finally, for large electric fields, the applied field can significantly amplify the velocity fluctuations and suppress fluctuations of salt concentration.  We expect this phenomenon to be observable experimentally and by molecular dynamics simulations. 

The theory developed here can readily be extended in a number of important directions. Firstly, the assumption of dynamically-identical ions can be removed so that a more direct comparison with experimental measurements for different salts can be performed, including polyvalent salts. It is also important to consider solutions with one ion and two counterions, such as for example solutions of NaCl and KCl in water. Such extensions would reveal whether the surprising experimental observation of negative Maxwell-Stefan diffusion coefficients\cite{NegativeMaxStefanDiff1, NegativeMaxStefanDiff2} between co-ions \cite{Electrolytes_DH_review} can be explained by fluctuating hydrodynamics and renormalization. Here we only considered strong electrolytes but the generalization to weak electrolytes is possible by using FHD for reactive fluids \cite{FluctReactDiff}. Lastly, we started here with fluctuating hydrodynamics equations based on the PNP equations, i.e., we assumed an ideal solution with no cross-diffusion, so our starting equations had only one mobility coefficient per ion, instead of one Maxwell-Stefan coefficient per pair of ions. The renormalized equations, on the other hand, have cross-diffusion and also a non-ideal Debye-Huckel contribution to the free energy density. This suggests that a more proper theory should start from the more complete equations, allowing for a nonzero \emph{bare} MS cross-coefficient $\mathfrak{D}^{(+,-)}_0$. As explained in \cite{CoarseBlob} for non-electrolytes, bare diffusion coefficients can be given a microscopic interpretation in terms of Green-Kubo expressions and can therefore, in principle, be measured in molecular dynamics simulations, and the renormalization due to thermal fluctuations computed numerically using a numerical fluctuating hydrodynamics solver \cite{LowMachElectrolytes}. Carrying out such an ambitious program for electrolyte solutions is a worthy challenge for the future.

\subsection*{Acknowledgements}

We thank Burkhard Duenweg and Mike Cates for illuminating discussions about linear response theory and renormalization.
This work was supported by the
U.S.~Department of Energy, Office of Science,
Office of Advanced Scientific Computing Research,
Applied Mathematics Program under Award Number DE-SC0008271 and contract DE-AC02-05CH11231.
This research used resources of the National Energy Research Scientific Computing Center, a DOE Office of Science User Facility supported by the Office of Science of the U.S. Department of Energy under Contract No. DE-AC02-05CH11231.

\bibliographystyle{ieeetr}
\bibliography{References,References_JP}

\begin{thebibliography}{10}

\bibitem{grodzinsky2011field}
A.~Grodzinsky, {\em Field, Forces and Flows in Biological Systems}.
\newblock Taylor \& Francis Group, 2011.

\bibitem{LowMachElectrolytes}
J.-P. P\'eraud, A.~Nonaka, A.~Chaudhri, J.~B. Bell, A.~Donev, and A.~L. Garcia,
  ``Low mach number fluctuating hydrodynamics for electrolytes,'' {\em Phys.
  Rev. Fluids}, vol.~1, p.~074103, 2016.

\bibitem{ExtraDiffusion_Vailati}
D.~Brogioli and A.~Vailati, ``{Diffusive mass transfer by nonequilibrium
  fluctuations: Fick's law revisited},'' {\em Phys. Rev. E}, vol.~63, no.~1,
  p.~12105, 2000.

\bibitem{DiffusionRenormalization_PRL}
A.~Donev, A.~L. Garcia, A.~de~la Fuente, and J.~B. Bell, ``{Diffusive Transport
  by Thermal Velocity Fluctuations},'' {\em Phys. Rev. Lett.}, vol.~106,
  no.~20, p.~204501, 2011.

\bibitem{DiffusionRenormalization}
A.~Donev, A.~L. Garcia, A.~de~la Fuente, and J.~B. Bell, ``{Enhancement of
  Diffusive Transport by Nonequilibrium Thermal Fluctuations},'' {\em J. of
  Statistical Mechanics: Theory and Experiment}, vol.~2011, p.~P06014, 2011.

\bibitem{GiantFluctuations_Nature}
A.~Vailati and M.~Giglio, ``{Giant fluctuations in a free diffusion process},''
  {\em Nature}, vol.~390, no.~6657, pp.~262--265, 1997.

\bibitem{FluctHydroNonEq_Book}
J.~M.~O.~D. Zarate and J.~V. Sengers, {\em {Hydrodynamic fluctuations in fluids
  and fluid mixtures}}.
\newblock Elsevier Science Ltd, 2006.

\bibitem{LowMachExplicit}
A.~Donev, A.~J. Nonaka, Y.~Sun, T.~G. Fai, A.~L. Garcia, and J.~B. Bell, ``{Low
  Mach Number Fluctuating Hydrodynamics of Diffusively Mixing Fluids},'' {\em
  Communications in Applied Mathematics and Computational Science}, vol.~9,
  no.~1, pp.~47--105, 2014.

\bibitem{FluctuationDissipation_Kubo}
R.~Kubo, ``{The fluctuation-dissipation theorem},'' {\em Reports on Progress in
  Physics}, vol.~29, no.~1, pp.~255--284, 1966.

\bibitem{robinson2012electrolyte}
R.~A. Robinson and R.~H. Stokes, {\em Electrolyte Solutions: Second Revised
  Edition}.
\newblock Dover Books on Chemistry Series, Dover Publications, Incorporated,
  2012.

\bibitem{Onsager1927}
L.~Onsager, ``Zur theorie der elektrolyte. ii,'' {\em Phys. Z.}, vol.~28,
  pp.~277--298, 1927.

\bibitem{ElectrolytesMS_Review}
J.~Wesselingh, P.~Vonk, and G.~Kraaijeveld, ``Exploring the maxwell-stefan
  description of ion exchange,'' {\em The Chemical Engineering Journal and The
  Biochemical Engineering Journal}, vol.~57, no.~2, pp.~75--89, 1995.

\bibitem{DiffusionJSTAT}
A.~Donev, T.~G. Fai, and E.~Vanden-Eijnden, ``{A reversible mesoscopic model of
  diffusion in liquids: from giant fluctuations to Fick's law},'' {\em Journal
  of Statistical Mechanics: Theory and Experiment}, vol.~2014, no.~4,
  p.~P04004, 2014.

\bibitem{IrrevThermoBook_Kuiken}
G.~D.~C. Kuiken, {\em Thermodynamics of Irreversible Processes: Applications to
  Diffusion and Rheology}.
\newblock Wiley, 1994.

\bibitem{Visser_Thesis}
C.~R. Visser, {\em Electrodialytic Recovery of Acids and Bases}.
\newblock PhD thesis, Rijksuniversiteit Groningen, Groningen, Netherlands,
  2001.
\newblock Available at
  \url{http://www.rug.nl/research/portal/files/14524647/thesis.pdf}.

\bibitem{LatticeBoltzElectrolytes}
J.~Zudrop, S.~Roller, and P.~Asinari, ``{Lattice Boltzmann scheme for
  electrolytes by an extended Maxwell-Stefan approach},'' {\em Phys. Rev. E},
  vol.~89, p.~053310, 2014.

\bibitem{Verpoorte2002}
E.~Verpoorte, ``Microfluidic chips for clinical and forensic analysis,'' {\em
  ELECTROPHORESIS}, vol.~23, pp.~677--712, 2002.

\bibitem{NegativeMaxStefanDiff1}
G.~Kraaijeveld and J.~A. Wesselingh, ``{Negative Maxwell-Stefan diffusion
  coefficients},'' {\em Industrial \& Engineering Chemistry Research}, vol.~32,
  no.~4, pp.~738--742, 1993.

\bibitem{NegativeMaxStefanDiff2}
G.~Kraaijeveld, J.~A. Wesselingh, and G.~D.~C. Kuiken, ``{Comments on "Negative
  Maxwell-Stefan Diffusion Coefficients"},'' {\em Industrial \& Engineering
  Chemistry Research}, vol.~33, no.~3, pp.~750--751, 1994.

\bibitem{Electrolytes_DH_review}
L.~M. Varela, M.~Garcia, and V.~Mosquera, ``Exact mean-field theory of ionic
  solutions: non-debye screening,'' {\em Physics reports}, vol.~382, no.~1,
  pp.~1--111, 2003.

\bibitem{FluctReactDiff}
C.~Kim, A.~J. Nonaka, A.~L. Garcia, J.~B. Bell, and A.~Donev, ``{Stochastic
  simulation of reaction-diffusion systems: A fluctuating-hydrodynamics
  approach},'' {\em J. Chem. Phys.}, vol.~146, no.~12, 2017.
\newblock Software available at
  \url{https://github.com/BoxLib-Codes/FHD_ReactDiff}.

\bibitem{CoarseBlob}
P.~Español and A.~Donev, ``{Coupling a nano-particle with isothermal
  fluctuating hydrodynamics: Coarse-graining from microscopic to mesoscopic
  dynamics},'' {\em J. Chem. Phys.}, vol.~143, no.~23, 2015.

\end{thebibliography}

\end{document}